\def\cm2{cm$^{-2}$}

\def\c2{C~{\sc ii}}
\def\c4{C~{\sc iv}}
\def\fe2{Fe~{\sc ii}}
\def\fe3{Fe~{\sc iii}}
\def\mg1{Mg~{\sc i}}
\def\mg2{Mg~{\sc ii}}
\def\si2{Si~{\sc ii}}
\def\si4{Si~{\sc iv}}
\def\al2{Al~{\sc ii}}
\def\al3{Al~{\sc iii}}
\def\o1{O~{\sc i}}
\def\n1{N~{\sc i}}
\def\h1{H~{\sc i}}

\def\approxlt{\mathrel{\spose{\lower 3pt\hbox{$\sim$}}
        \raise 2.0pt\hbox{$<$}}}
\def\approxgt{\mathrel{\spose{\lower 3pt\hbox{$\sim$}}
        \raise 2.0pt\hbox{$>$}}}

\documentclass[article]{gwp80}
\usepackage{graphicx}
\usepackage{amssymb}
\tabletypesize{\normalsize}  

\shortauthors{Sylvie Vauclair}
\shorttitle{Element Diffusion and Accretion in Metal Poor Stars}

\begin{document}
\large    
\pagenumbering{arabic}
\setcounter{page}{216}

\title{Element Diffusion and Accretion in Metal Poor \\ \\ Stars}

%
%
\author{\noindent Sylvie Vauclair{$^{\rm 1,2}$}
\\
{\it (1) Institut de Recherches en Astrophysique et Plan\'etologie, CNRS, Universit\'e Paul Sabatier, Toulouse, France\\
(2) Institut universitaire de France} 
}

%
%
\email{(1) sylvie.vauclair@ast.obs-mip.fr}


\begin{abstract}

The abundances of the chemical elements observed at the surface of metal-poor 
stars are not always representative of their initial values. 
During stellar evolution, various physical processes modify their internal composition.
In this short paper, in honor of George W. Preston, I remind the importance of atomic diffusion, 
and the possible effects of
accretion processes which may lead to double-diffusive (thermohaline) instabilities. 
I discuss the consequences of these processes and compare them to current observations.

\end{abstract}

\section{Introduction}

The chemical abundances observed in metal poor stars are often considered as a clue 
for the understanding of the chemical evolution of galaxies, and they are generally assumed to 
be representative of their initial values. This assumption does not take into account the basics 
of stellar physics, which lead to the conclusion that the element abundances must change with
time during stellar evolution, due to atomic diffusion acting alone or combined with the various macroscopic
processes which may occur inside the stars. 

An interesting case is that of lithium in halo stars. Stellar physicists 
claimed for a long time that the abundances of lithium in these stars cannot be the original ones, the computations
showing that it has decreased by at least a factor two since the origin, due to atomic diffusion
coupled with nuclear reactions (Michaud et al. \cite{michaud84}, Vauclair \cite{vauclair88}, Vauclair \& Charbonnel \cite{vauclair95} and \cite{vauclair98}). In spite of these results,
the observed lithium value was still presented by the observers as the primordial one, and used as a constraint for the
Big Bang nucleosynthesis. Later on, the space mission WMAP proved that the baryonic
density obtained from the cosmological background is different from that obtained 
with the observed lithium value, and that it is compatible with a smaller
primordial value, as predicted by stellar physicists.

Atomic diffusion is more complex to compute for heavy elements than for
light ones, because of their more complicated atomic structure. It should, however, always
been taken into account in the interpretation of the observations. In the following, I first remind the basics of atomic diffusion,
and I give examples of its importance for metal-poor stars (Section 2). In section 3, I discuss
another process acting on some metal poor stars: accretion followed by thermohaline convection. These processes
are connected, as the abundance gradients induced by atomic diffusion play an important role
in the extent of the mixing. I give my conclusions in Section 4. Without George Preston, I would not have
been involved in the problem of accretion onto metal poor stars, which is a very interesting subject from 
both observational and physical view points. Much thanks to you, George, on your birthday.

\section{Atomic diffusion in stars}

The fundamental importance of atomic diffusion inside stars was recognized
by the pioneers of stellar physics like Eddington \cite{eddington16}. Chapman \cite{chapman17} already discussed 
the ``relative importance of convection and diffusion within a star''. At that time, he
only introduced in the computations the effects of gravitational and thermal diffusion.
In this case, he found that no heavy elements should be left at the surface of stars,
which was obviously wrong.

A first approach of the possible effects of selective radiation pressure, pushing the
elements upwards in a way related to their atomic structure, may be found in Eddington \cite{eddington26}.
He explained how the radiation force is much stronger on heavy atoms than on light ones because 
of line absorption: the heavy elements should be pushed up at the surface of stars, with an effect
increasing with increasing stellar mass. But this was not observed at that time and was put at the
back of the mind of astronomers for a long time. 

Aller \& Chapman \cite{aller60} came back to this problem and computed atomic diffusion time scales for metals 
at the bottom of the solar convective zone. They found that this time scale was longer than the
solar age, so that there should be no important effect of atomic diffusion inside the Sun.

Some years later, the presence of chemical peculiarities in main sequence stars became more and more evident,
and the explanation in terms of atomic diffusion more and more successful (see references in Vauclair \& Vauclair (\cite{vauclair82}).
It was still believed however that atomic diffusion occurred only 
in specific stars, and was suppressed by convection or other macroscopic motions in most of the other stars. In particular, 
the small effect of helium and metal diffusion in the Sun was considered as negligible, 
until helioseismology proved in a spectacular way that it really occurs as predicted and has a measurable influence on 
the internal solar sound velocity (Gough et al. \cite{gough96}, Richard et al. \cite{richard96}).

From then on, atomic diffusion in stars has been much studied. The most difficult problem in the computations 
is the evaluation of selective radiative accelerations on each ion of each element. This is related to the computations 
of opacities, as 
radiative accelerations act through the absorption of the radiation flux by the ions. At the present time, three stellar
evolution codes are able to compute atomic diffusion including radiative accelerations: the Montreal code (C.f. Richard et al. \cite{richard02}), the TGEC code
(Toulouse Geneva Evolution Code, C.f. Theado et al. \cite{theado10}) and the Yale code (Delahaye \& Pinsonneault \cite{delahaye05}).

Atomic diffusion cannot occur inside convective zones, due to the efficient mixing, and it is subject to competition with extra-mixing processes inside radiative zones. These processes can be due to rotation-induced mixing, internal waves, mass loss, thermohaline convection, etc. They have been extensively discussed in the recent literature (see Th\'eado \& Vauclair \cite{theado03}, Th\'eado et al. \cite{theado10}, Talon \& Charbonnel \cite{talon05} and \cite{talon08}, Dennissenkov et al.\cite{denissenkov10}, etc.).  Due to all these macroscopic processes, atomic diffusion is slowed down and the final results modified, but it is generally not completely suppressed. Atomic diffusion alone is parameter free, which is not the case for mixing. The detailed observation of the abundances of many elements can lead to strong constraints on these macroscopic processes.

Atomic diffusion in metal poor stars is important for various reasons. 

\begin{itemize}

\item The helium settling leads to modifications of the stellar evolution tracks, 
and this may change the evaluated age of globular clusters. Van den Berg et al.\cite{berg02} have found that
introducing atomic diffusion can reduce the age by about ten percent.

\item The presence of diffusion-induced helium gradients below the convective zones can have important secondary effects,
for example in the case of metal-rich accretion, as discussed below (Section 3).

\item Beyond the lithium problem, discussed in the introduction, atomic diffusion modifies the detailed abundances of heavy elements in halo stars (e.g. Richard et al. \cite{richard02} and \cite{richard05}, Korn et al. \cite{korn06} and \cite{korn09}). Concerning lithium, diffusion alone would lead to results incompatible with the observations, as it should be more deficient in more massive stars, which is not observed. Mixing processes are needed to slow down the selective process. However, such a mixing connects the bottom of the convective zone to the lithium nuclear destruction layers, so that in any case lithium is depleted. The fact that lithium is not completely absent from the spectra of metal deficient stars leads to constraints on the mixing processes which must not be too strong. As a consequence, atomic diffusion still plays a role and this is reflected on the abundances of metals (e.g. discussions in Vauclair \& Charbonnel \cite{vauclair98} and Korn et al. \cite {korn06})

\end{itemize}

\section{Accretion and thermohaline convection}

The abundances at the surface of stars may sometimes be altered by the accretion of external matter. Two interesting cases 
have recently been studied. The first case concerns the accretion of planetary matter onto exoplanet-host stars. Vauclair \cite{vauclair04} 
and more recently Th\'eado et al. \cite{theado10} and Th\'eado \& Vauclair \cite{theado11}, and Garaud \cite{garaud11} have discussed 
how the accreted matter is mixed down due
to double-diffusive (thermohaline) convection induced by the inverse $\mu$-gradient.

Here I will discuss another case, in which George Preston is particularly interested: that of Carbon Enriched Metal Poor Stars (CEMP).

CEMP stars are metal poor stars
([Fe/H] smaller than -2), with a large overabundance of carbon with respect to the other heavy elements ([C/Fe] larger than +1.0). The number of these stars increase
with decreasing metallicity (Lucatello et al. \cite{lucatello06}).
Most of them are rich in s-
process elements, forming the
so-called CEMP-s group. 
The commonly accepted explanation
of the C and s process elements enrichment is these stars is that they were submitted
to a shower of 
matter coming from an AGB companion which later on can have become a white dwarf.

A complication comes from the fact that about half of the CEMP-s stars also show a high r-process
enrichment, incompatible with a pure s-process contribution. 
The nucleosynthesis of the r-process does not come from AGBs, but it is believed to come from massive 
stars exploding as Type II Supernovae. An interpretation of these observations is that the r-process enrichment was already present in the
molecular cloud from which the binary system formed (Bisterzo et al. \cite{bisterzo10}). 

Stancliffe et al. \cite{stancliffe07} rejected the accretion hypothesis on the basis that in the case of accretion of metal-rich matter, thermohaline convection develops so that the accreted matter is rapidly diluted throughout the star. Thompson et al. \cite{thompson08} pointed out that the main sequence star which suffers accretion was subject to helium settling since its birth so that it could have built a stable $\mu$-gradient in its outer layers. Such a gravitational-settling induced $\mu$-gradient can stop at least some of the accretion-induced thermohaline mixing. In this case the chemical composition of the stellar atmosphere can be modified by the accreted material. Later on, Stancliffe et al. \cite{stancliffe08} performed new computations including gravitational settling and confirmed that it may slow down or even inhibit the thermohaline mixing. In their computations, they used the prescription suggested by Charbonnel \& Zahn \cite{charbonnel07} for Red Giants, which is now believed to be overestimated. We can thus infer that the mixing effect in CEMPs is still smaller than previously thought, so that the accretion-induced modification of the abundances may be larger than previously thought. Let me discuss this in more detail

Thermohaline convection is a well known process in oceanography : warm
salted layers on the top of cool unsalted ones rapidly diffuse downwards even
in the presence of stabilizing temperature gradients, due to the different diffusivities of heat and salt. 
When a warm salted blob falls down in cool fresh water, the heat diffuses out more quickly than the salt. 
The blob goes on falling due to its weight until it mixes with the surroundings. 
This leads to the so-called ``salt fingers''. Thermohaline convection is a ``double diffusive convection''.  
When the gradients are reversed, another kind of double diffusive convection occurs, which is generally referred to as ``semi convection''. 
The condition for the salt fingers to develop is related to the
density variations induced by temperature and salinity perturbations.

Thermohaline convection occurs in stellar radiative zones in the presence of inverse $\mu$-gradients. 
The medium
can become dynamically unstable if (Ledoux criterion):
\begin{equation}
\nabla_{crit} = \frac{\phi}{\delta}\nabla_{\mu} + \nabla_{ad} - \nabla < 0  
\end{equation}
where $\phi=(\partial$ ln $\rho/\partial$ ln $\mu)$ and $\delta=(\partial$ ln $\rho/\partial$ ln $T)$.
When $\nabla_{crit}$ vanishes, marginal stability is achieved and thermohaline convection may begin as a ``secular process", 
namely on a thermal time scale (short compared to the stellar lifetime).

The effects of thermohaline convection as a mixing process in stars are far from trivial. 
First parametrization recipes were given by Ulrich \cite{ulrich72} and Kippenhahn et al. \cite{kippenhahn80}. 
These authors treated thermohaline mixing as a diffusion process with a diffusion coefficient depending on the relative size of the ``fingers'', and more specifically on the ratio of their horizontal to vertical scale, which was totally unknown. Furthermore, a multiplying coefficient was introduced, of order 12 for Kippenhahn et al. \cite{kippenhahn80}, and much larger for Ulrich \cite{ulrich72}. Charbonnel \& Zahn \cite{charbonnel07} found that, to account for the abundances observed in red giants, this coefficient should be of order 1000. 

More recently, numerical simulations of thermohaline convection in conditions closer to the stellar ones than before have been performed by Traxler et al. \cite{traxler11} and Garaud \cite{garaud11}. They derive a new diffusion coefficient which is of the same order of magnitude as found from the Kippenhahn et al. \cite{kippenhahn80} prescription, much below that suggested by Charbonnel \& Zahn \cite{charbonnel07}. This result is confirmed through different numerical computations by Denissenkov \& Merryfield \cite{denissenkov11}. It is also shown by Th\'eado \& Vauclair \cite{theado11} that the large mixing coefficient needed for the Red Giants would lead to more lithium depletion than observed in case of planetary material accretion onto the star.

\section{Conclusion}

I first met George W. Preston in 1978, when, during a long stay in Cal Tech, I had the privilege to do some observations with my husband G\'erard Vauclair at the 200 inch telescope in Mount Wilson. We did spectroscopic observations of Am stars, and George was our (extremely nice and efficient) support astronomer. We had afterward some festive evenings, doing music together (I still have the american folksong book he gave me at that time and I use it for my grand daughters!). At that time, George already told us about metal-poor stars, but we were more involved with metal-rich ones...

Time has passed, and it was a great pleasure for me to be sought by him a few years ago about the efficiency of thermohaline convection in CEMPs stars, in case of AGB accretion. I remember he really wanted me to discuss about mixing coefficients and time scales. I thought about it and then came in my mind the idea of the helium-induced stabilizing $\mu$-gradient. Such a situation does not occur for the accretion of planetary matter onto exoplanet-host stars, which I studied before, because the stars are very young when they suffer the metal-rich accretion. But when old main-sequence stars accrete AGB matter, they obviously had time to built such a gradient. Helioseismology has proved that helium settling is at work, even in stars with deep convective zone as in the Sun.

This was for me a very interesting subject, which I would not have entered without George. The story is not finished, and much work still has to be done on the simulations of various situations were accretion, thermohaline mixing and gravitational settling work together, besides nuclear reactions, to lead to the rich variety of observed element abundances. Nature is complex and fascinating, and we are lucky to work on a better understanding of all this complexity. Thanks and long live!

\end{document}